\documentclass[12pt,preprint]{aastex}

\slugcomment{}

\shorttitle{MULTI-TeV GAMMA-RAY FLARES FROM MARKARIAN~421}
\shortauthors{Amenomori et al.}

\begin{document}

\title{MULTI-TeV GAMMA-RAY FLARES \\ 
FROM MARKARIAN~421 IN 2000 AND 2001 OBSERVED WITH THE TIBET AIR SHOWER ARRAY
}

\author{ M.~Amenomori\altaffilmark{1},  S.~Ayabe\altaffilmark{2},
	S.W.~Cui\altaffilmark{3},       Danzengluobu\altaffilmark{4},
	L.K.~Ding\altaffilmark{3},      X.H.~Ding\altaffilmark{4},      
	C.F.~Feng\altaffilmark{5},      Z.Y.~Feng\altaffilmark{6},      
        X.Y.~Gao\altaffilmark{7},       Q.X.~Geng\altaffilmark{7}, 
        H.W.~Guo\altaffilmark{4},       H.H.~He\altaffilmark{3},
	M.~He\altaffilmark{5},          K.~Hibino\altaffilmark{8},
	N.~Hotta\altaffilmark{9},  	Haibing Hu\altaffilmark{4},
	H.B.~Hu\altaffilmark{3},        J.~Huang\altaffilmark{9},
	Q.~Huang\altaffilmark{6},	H.Y.~Jia\altaffilmark{6},       
	F.~Kajino\altaffilmark{10},	K.~Kasahara\altaffilmark{11},
	Y.~Katayose\altaffilmark{12},   K.~Kawata\altaffilmark{10},
	Labaciren\altaffilmark{4},	G.M.~Le\altaffilmark{13},
	J.Y.~Li\altaffilmark{5},        H.~Lu\altaffilmark{3},
	S.L.~Lu\altaffilmark{3},	X.R.~Meng\altaffilmark{4}, 
        K.~Mizutani\altaffilmark{2},    J.~Mu\altaffilmark{7}, 
        H.~Nanjo\altaffilmark{1},       M.~Nishizawa\altaffilmark{14},
        M.~Ohnishi\altaffilmark{15},    I.~Ohta\altaffilmark{9},
        T.~Ouchi\altaffilmark{15},      S.~Ozawa\altaffilmark{9},
	J.R.~Ren\altaffilmark{3},       T.~Saito\altaffilmark{16},
	M.~Sakata\altaffilmark{10},     T.~Sasaki\altaffilmark{8},
	M.~Shibata\altaffilmark{12},    A.~Shiomi\altaffilmark{15},
	T.~Shirai\altaffilmark{8},      H.~Sugimoto\altaffilmark{17},
	K.~Taira\altaffilmark{17},      M.~Takita\altaffilmark{15},
	Y.H.~Tan\altaffilmark{3},       N.~Tateyama\altaffilmark{8},
	S.~Torii\altaffilmark{8},       H.~Tsuchiya\altaffilmark{15},
	S.~Udo\altaffilmark{2},		T.~Utsugi\altaffilmark{8},
	B.S.~Wang\altaffilmark{3},	H.~Wang\altaffilmark{3},
	X.~Wang\altaffilmark{2},	Y.G.~Wang\altaffilmark{5},
	L.~Xue\altaffilmark{5},		Y.~Yamamoto\altaffilmark{10},
	X.C.~Yang\altaffilmark{7},	Z.H.~Ye\altaffilmark{13},
	G.C.~Yu\altaffilmark{6},	A.F.~Yuan\altaffilmark{4},
	T.~Yuda\altaffilmark{15,18},	H.M.~Zhang\altaffilmark{3},
	J.L.~Zhang\altaffilmark{3},	N.J.~Zhang\altaffilmark{5},
	X.Y.~Zhang\altaffilmark{5},	Y.~Zhang\altaffilmark{3},
	Zhaxisangzhu\altaffilmark{4},   and X.X.~Zhou\altaffilmark{6}\\
(The Tibet AS${\bf \gamma}$ Collaboration)}

\altaffiltext{1}{  Department of Physics, Hirosaki University, Hirosaki 036-8561, Japan}
\altaffiltext{2}{  Department of Physics, Saitama University, Saitama 338-8570, Japan}
\altaffiltext{3}{  Institute of High Energy Physics, Chinese Academy of Sciences, Beijing 100039, China}
\altaffiltext{4}{  Department of Mathematics and Physics, Tibet University, Lhasa 850000, China}
\altaffiltext{5}{  Department of Physics, Shandong University, Jinan 250100, China}
\altaffiltext{6}{  Institute of Modern Physics, South West Jiaotong University, Chengdu 610031, China} 
\altaffiltext{7}{  Department of Physics, Yunnan University, Kunming 650091, China}
\altaffiltext{8}{  Faculty of Engineering, Kanagawa University, Yokohama 221-8686, Japan}
\altaffiltext{9}{  Faculty of Education, Utsunomiya University, Utsunomiya 321-8505, Japan}
\altaffiltext{10}{ Department of Physics, Konan University, Kobe 658-8501, Japan}
\altaffiltext{11}{ Faculty of Systems Engineering, Shibaura Institute of Technology, Saitama 330-8570, Japan} 
\altaffiltext{12}{ Faculty of Engineering, Yokohama National University, Yokohama 240-8501, Japan}
\altaffiltext{13}{ Center of Space Science and Application Research, Chinese Academy of Sciences, Beijing 100080, China}
\altaffiltext{14}{ National Institute for Informatics, Tokyo 101-8430, Japan}
\altaffiltext{15}{ Institute for Cosmic Ray Research, the University of Tokyo, Kashiwa 277-8582, Japan}
\altaffiltext{16}{ Tokyo Metropolitan College of Aeronautical Engineering, Tokyo 116-0003, Japan}
\altaffiltext{17}{ Shonan Institute of Technology, Fujisawa 251-8511, Japan}
\altaffiltext{18}{ Solar-Terrestrial Environment Laboratory, Nagoya University, Nagoya 464-8601, Japan}

\begin{abstract}

Several strong TeV $\gamma$-ray flares were detected from
Markarian~421 in the years 2000 and 2001 by the Tibet~III air shower
array at a level of statistical significance of 5.1~$\sigma$.
Markarian~421 was unprecedentedly active at X-ray and TeV $\gamma$-ray
energies during this period, and a positive correlation was found
between the change of the ASM/RXTE X-ray flux and the Tibet TeV
$\gamma$-ray flux. When a power-law energy spectrum for
$\gamma$-rays from this source is assumed, the spectral index is
calculated to be $-$3.24$\pm$0.69 at the most active phase in 2001.
The spectral index observed by the Tibet air shower
array is consistent with those obtained via imaging air Cerenkov
telescopes.
\end{abstract}

\keywords{BL Lacertae objects : individual (Markarian~421) -- gamma rays : observations}

\section{INTRODUCTION}

A variable $\gamma$-ray source Markarian~421 (Mrk~421) at redshift
$z~=~0.031$ is known to be a blazar class of active galactic nuclei
(AGNs) with a common feature of BL Lac objects.  Since a relativistic
jet from this AGN is aligned along our line of sight, its photon
emission is dominated by the jet power output, which is mainly
non-thermal, extending over more than fifteen energy decades.  In
1991, the EGRET detected $\gamma$-ray emissions from Mrk~421, which was
the first detection of a BL Lac-type $\gamma$-ray source.  Its
integrated photon flux above 100 MeV was $1.4\pm0.3 \times 10^{-7}$
cm$^{-2}$ s$^{-1}$, and its differential photon energy spectrum can be
represented by a power law with an exponent of $-$1.96$\pm$0.14
\cite{Lin1992}.  Subsequently, the first detection of TeV
$\gamma$-rays from this source was made by the Whipple collaboration
(Whipple) in 1992.  The average integral flux was estimated to be $1.5
\times 10^{-11}$ cm$^{-2}$ s$^{-1}$ above 0.5 TeV, being 0.3 times as
large as that from the Crab Nebula which often serves as the standard
candle in TeV $\gamma$-ray astronomy \cite{Punch1992}.

Occasionally, the $\gamma$-ray flux from Mrk~421 shows a rapid
variability at TeV energies.  The Whipple observed significant
variabilities on a few-hour time scale on May 14 and 15 in 1994,
during which the average source flux above 250 GeV increased by a
factor of $\sim$10 \cite{Kerrick1995}.  Such rapid variabilities of
TeV $\gamma$-ray flux were detected again by the Whipple during the
period from April 20 through May 5 in 1995 \cite{Buckley1996} and in
May, 1996 \cite{Gaidos1996} at sub-TeV energies.  At the second
outburst in 1996, which lasted for about 30 minutes, the flux
increased by a factor of 20-25.

Various theoretical models of photon emission mechanisms are
based basically on the Synchrotron self-Compton (SSC) model \cite{Gould1965},
combining the synchrotron photons and inverse Compton (IC) photons
produced by accelerated high-energy electrons which interact with the
magnetic field and the synchrotron photons, respectively, in a jet.
This model naturally leads to a photon energy spectrum having two
broad continuous peaks, one at energies between the infrared and X-ray
regions and the other at energies between the GeV and TeV regions.  This
lower-energy peak is attributed to synchrotron radiation from
accelerated high-energy electrons in the AGN, and the higher-energy
peak is due to inverse Compton scattering of the same electrons off
the soft photons \citep{Maraschi1992,Marscher1996}.  In 1995 and 1998,
a simultaneous multi-wavelength observation of Mrk~421 was made,
covering the radio, optical, and X-ray bands as well as the MeV-TeV
$\gamma$-ray regions.  It was shown from this observation that the
multi-wavelength spectrum observed is consistent with that expected
from the SSC model, suggesting a possible coincidence of X-ray and TeV
$\gamma$-ray emissions \citep{Macomb1995,Takahashi1996,Takahashi2000}.

It has been suggested that TeV $\gamma$-rays from nearby AGNs are
absorbed rapidly due to their interaction with the infrared photon
field in the intergalactic space.  Therefore, the shape of the multi-TeV
$\gamma$-ray energy spectrum provides us important information on the
column density and energy spectrum of intergalactic infrared
photons.  From this point of view, TeV $\gamma$-ray data have been
used to impose an upper limit of these photons that are difficult to
measure directly \citep{DeJager1994,Biller1995}.

In the years 2000 and 2001, fortunately, Mrk~421 entered a very
active phase, showing strong and frequent flaring of X-rays and TeV
$\gamma$-rays.  During this period, TeV $\gamma$-rays from Mrk~421
were detected via various imaging air Cerenkov telescopes (IACTs)
used in the Whipple \cite{Krennrich2001}, the HEGRA
\cite{Krowczynski2001}, the CAT \cite{Piron2001}, and the CANGAROO 
\cite{Okumura2002} experiments. The Tibet air shower experiment also
successfully detected multi-TeV $\gamma$-rays from this source.

In this paper, we report the result on the flux of $\gamma$-rays from
Mrk~421 observed by this array during its outbursting period in the
years 2000 and 2001 in comparison with those from the ASM/RXTE satellite
and IACTs.

\section{EXPERIMENT}

The Tibet air shower experiment has been successfully operated at
Yangbajing ($90.^{\circ}522$E, $30.^{\circ}102$N; 4,300~m above sea
level) in Tibet, China since 1990. The Tibet I array constructed in
1990 \cite{Amenomori1992} was gradually updated by increasing the
number of counters in 1995 and 1996, as briefly described elsewhere
\cite{Amenomori2000, Amenomori2002}.  Using this array, we
successfully observed multi-TeV $\gamma$-ray signals from the Crab~Nebula 
in 1999 \cite{Amenomori1999} and multi-TeV $\gamma$-ray flares
from Mrk~501 in 1997 \cite{Amenomori2000}.

In the late fall of 1999, the array was further updated by adding 235
scintillation counters so as to enlarge a high density array,
Tibet~III, with a 7.5~m lattice interval as shown in Figure~1.  The
Tibet~III array consists of 533 scintillation counters covering 22,050
m$^2$. Each counter has a plastic scintillator plate (BICRON BC-408A)
of 0.5~m$^{2}$ in area and 3 cm in thickness and is equipped with a
fast timing (FT) photomultiplier tube (PMT, Hamamatsu H1161).  A
0.5~cm-thick lead plate is put on the top of each counter in order to increase
the array sensitivity by converting $\gamma$-rays into
electron-positron pairs in the shower
\citep{Bloomer1988,Amenomori1990}. The relative position of each
counter is measured by the use of a portable GPS (Global Positioning
System) with an inaccuracy of less than 2 cm.  Timing and charge
information from each hit PMT is digitized by the use of a
time-to-digital converter (TDC, LeCroy 1887) and a charge-sensitive
analog-to-digital converter (ADC, LeCroy 1881M), respectively.  All
the TDCs and ADCs are regularly monitored using a FASTBUS calibration
module (CAT, LeCroy 1810) every 20 minutes.  The length of each signal
cable is also monitored, with an inaccuracy of 0.1 ns, by measuring a
mismatched-reflection pulse from each counter every 20 minutes.

An event trigger signal is issued when an any-four coincidence appears in
the FT counters each recording more than 0.6 particles within a
coincidence gate width of 300 ns, resulting in the trigger rate of
about 680~Hz. All triggered data are collected using a computer via
FASTBUS-to-VMEbus interface and stored on the data tapes
($\sim$20GB/day).  We collected 2.7$\times$10$^{10}$ events during the
period from November 17, 1999 through October 10, 2001, and the
live time was calculated to be 456.8 days.  The event
selection was made by imposing the following conditions on the
recorded data: (1) each shower event should fire four or more
FT-counters recording 1.25 or more particles, (2) among the 9 hottest
FT-counters in each event, 8 should be contained in a fiducial area
enclosed by the dashed line in Figure~1.  If the number of hit counters
is less than 8, the numbers of all should be contained in its area, and (3)
the zenith angle of the arrival direction should be less than
40$^{\circ}$.  After this data selection and quality cuts,
5.52$\times$10$^{9}$ events remain for further analysis. The mode
energy of air shower events, thus obtained, is estimated to be about 3
TeV \cite{Amenomori1999}, covering the upper part of the energies
measured by the atmospheric Cerenkov technique.

\section{PERFORMANCE OF THE TIBET III ARRAY}

In order to successfully detect transient $\gamma$-ray signals from a
point source, it is important to check the stable operation of
equipment over a long period of time regarding such factors as variation of event
rate, pointing accuracy, and the angular resolution of the array.

The long-term stability of the daily event rate is shown in Figure~2,
where its variation is shown to be less than $\pm$5\% and it is mostly
due to atmospheric pressure and temperature effects.  The stability in
the pointing accuracy and angular resolution of the Tibet~III array
can be directly checked by monitoring the Moon's shadow in the cosmic
ray flux \cite{Amenomori2000}.  The statistical significance of the
Moon's shadow observed via the Tibet~III array is 26~$\sigma$ 
for 1.3-year observation, that is, a 7 $\sigma$ level per month.
The displacement of the center of the Moon's shadow in the north-south
direction enables us to estimate the systematic error in pointing
accuracy and its long-term stability, since the east-west component of
the geomagnetic field is almost zero at Yangbajing.  The displacements
of the shadow's center from the apparent center in the north-south
direction are plotted as a function of observation time in terms of
a sidereal month (27.3 days) in the upper panel of Figure~3.  From
this figure, it is estimated that the systematic pointing error per
month is smaller than $0.^{\circ}1$, while the overall pointing error
for the observation period is smaller than $0.^{\circ}02$.  The
amount of cosmic-ray deficits by the Moon provides a good measure of
the angular resolution. The lower panel in Figure~3 demonstrates the
long-term stability of the ratio of the observed deficit counts to the
expected ones within a window of the angular radius $0.^{\circ}9$.
The data show a variation smaller than 8\%, confirming the stability
of angular resolution.

The performance of the Tibet~III array is also examined by means of a full Monte
Carlo (MC) simulation. We used the Corsika Ver.~6.004 code
\cite{Corsika1998} for the generation of air shower events and the
Epics uv7.24 code \cite{Epics} for the detection of shower particles
with scintillation counters, respectively. Primary particles were
sampled from the energy spectrum made by using direct
observational data \citep{Asakimori1998,Sanuki2000,Kamioka2001} in the
energy range from 0.3 TeV to 1000 TeV. The absolute flux of the
primary particles is estimated with the error of about $\pm$20\% at the lower
energy region, while its uncertainty may become larger than $\pm$50\% at
energies higher than 100 TeV.  Figure~4 shows the size spectrum of
observed events, where the size is expressed by $\sum\rho_{\rm FT}$
being the sum of the number of particles per m$^2$ for each FT counter.
The observed spectrum shows close agreement with the simulation both
in terms of the event rate and the shape, as shown in Figure~4. After the
event reduction described in $\S$2, the event rates are 140~Hz for the
experiment and 130$\pm$1~Hz for the simulation, respectively.  For
more details regarding the performance of the Tibet~III array, see
Amenomori et al.\ \citep{Amenomori2001a,Amenomori2001b}.

\section{ANALYSIS}

In order to extract an excess of TeV $\gamma$-ray events coming from
the direction of Mrk~421, the background event density must be
carefully estimated.  The background is estimated by the number of
events averaged over 8 off-source cells with the same angular radius
as on-source, at the same zenith angle, recorded at the same time
intervals as the on-source cell events.  The search window radius is
expressed as $6.9/\sqrt{ \sum\rho_{\rm FT} }$ degrees as a function of
$\sum\rho_{\rm FT}$, which maximizes the $S/\sqrt{N}$ ratio according
to a MC study as shown in Figure~5.  This angular radius can be used
for the analysis of various $\gamma$-ray sources, while depending
weakly on the orbital motion of a source.  The center positions of
these off-source cells, located at every $3.^{\circ}2$ step from the
source position measured in terms of angle distance in the azimuthal
direction at the same zenith angle as the on-source direction, move
picking up events recorded in the common time interval to the
on-source cell.  Here, it is worthwhile to note that two 
off-source cells adjacent to the on-source cell are excluded in order to avoid a
possible signal tail leaking in the off-source events.  This method,
the so-called ``equi-zenith angle background estimation'', can reliably
estimate the background events under the same condition as on-source
events.  The Tibet~III array, however, has a small anisotropy of
$\pm$1.5\% in maximum amplitude in the azimuthal direction, as the
array is constructed on the ground with a slight slope of $+1.^\circ$3
to the normal plane in the northwest direction.  Hence, we analyzed 71
different dummy sources which follow the same diurnal rotation (at the
same declination) as that of Mrk~421 using the equi-zenith angle method, and
corrected the anisotropy of off-source events using the azimuthal
distribution averaged over 71 dummy sources events.  In the case of
Mrk~421, the number of off-source events decreases by
(0.16$\pm$0.02)\%.  It is noted that the equi-zenith angle method
fails when the source object stays very close to the zenith since an
off-source cell overlaps with other cells.  This method can be used
when a point source exists at zenith angles larger than
6$^{\circ}$. In the case of Mrk~421, its zenith angle at the
southing is about 8$^{\circ}$ at Yangbajing, so that each
on/off-source cell is independent throughout the observation period.

\section{RESULTS AND DISCUSSIONS}

We calculated the statistical significance of TeV $\gamma$-ray signals
from Mrk~421 using the following formula \citep{Li1983}: $(N_{\rm ON} -
\alpha N_{\rm OFF}) / \sqrt{\alpha (N_{\rm ON} + N_{\rm OFF})}$, where
$N_{\rm ON}$, $N_{\rm OFF}$, and $\alpha$ are the number of events in
the on-source cell, the number of background events summed over 8
off-source cells, and the ratio of on-source solid angle area to
off-source solid angle areas ($\alpha$ = 1/8 in this work),
respectively.

The number of events after the event reduction is plotted in Figure~6 as a
function of angular distance from Mrk~421. A clear peak of
$\gamma$-rays from Mrk~421 is seen at 5.1~$\sigma$ statistical
significance above the flat cosmic-ray background for the observation
of 456.8 live days.

\subsection{TeV $\gamma$-RAY/keV X-RAY CORRELATION}

We plot our daily excess event rate from Mrk~421 averaged over every
month in Figure~7, together with quick-look results from the all-sky
monitor on the Rossi X-Ray Timing Explorer (ASM/RXTE)
\citep{Levine1996,RXTEhtml} and those from three IACTs: the Whipple
\cite{Holder2001}, HEGRA \cite{Kohnle2001}, and CAT \cite{Piron2001}.
Since the Tibet~III array was operated very stably during this period
as discussed in $\S$3, the observed variation cannot be attributed to
any artificial noises or unstable data-acquisition system.
Furthermore, one can see an excellent correlation among the 5 experiments
as shown in Figure~7.  The Tibet~III array can observe TeV
$\gamma$-rays continuously regardless of day/night, while the
ASM/RXTE satellite orbits the Earth at a cycle of about
90 minutes, monitoring the X-rays continuously. 
Figure~8 shows the observation time overlap between the Tibet~III
array ($T_{\rm Tibet}$) and the ASM/RXTE satellite ($T_{\rm ASM}$).
It tells us that ($T_{\rm Tibet}$ $\cap$ $T_{\rm ASM}$) / $T_{\rm ASM}$ 
is 32.7\%, 
while ($T_{\rm Tibet}$ $\cap$ $T_{\rm ASM}$) / $T_{\rm Tibet}$ is 2.5\%.  
These constant overlapping fractions are
sufficient to study the TeV $\gamma$-ray/keV X-ray correlation with
both data sets.

Figure~9 shows a correlation plot of excess counts between the
ASM/RXTE data (2nd panel in Fig.\ 7) and the Tibet~III data (3rd
panel in Fig.\ 7) during the overlap period.  A positive
correlation is observed between the number of excess events in
Tibet~III ($N_{\rm Tibet}$) and those in ASM/RXTE ($N_{\rm ASM}$),
which is given by 
$N_{\rm Tibet}$ = (10.66$\pm$1.83) $N_{\rm ASM}$
with $\chi^{2} / d.o.f. = 17.6 / 19$ or
$N_{\rm Tibet}$ = (4.16$\pm$0.81) $N_{\rm ASM}^{2}$ 
with $\chi^{2} / d.o.f. = 24.6 / 19$, where d.o.f. is degree of freedom. 
This positive flux correlation between the keV and the TeV
regions will be consistent with the SSC model prediction. Furthermore,
the correlation between Tibet~III data and ASM/RXTE satellite data
appears linear rather than quadratic, although the
statistical significance is not sufficient,
which may suggest some
contribution of soft photons other than synchrotron photons to the TeV
$\gamma$-rays as discussed below.

A comparison between the multi-wavelength spectra of Mrk~421 and Mrk~501 was
made based on a 5-year observation from 1993 to 1998 via the ASCA
and the RXTE satellites \cite{Kataoka2001}.  According to this paper,
there is a conspicuous difference between Mrk~421 and Mrk~501 regarding the
synchrotron component of the photon spectrum.  In the case of Mrk~501,
the position of the peak energy in the synchrotron component shifts
manifestly from low energies to high energies as the source becomes
brighter, but that of Mrk~421 remains almost constant.  They
consider that the time variation of the flux intensity may be caused by
the increase of the number of high-energy electrons in the case of
Mrk~421, while it is caused by the increase of the maximum
acceleration energy of electrons in the case of Mrk~501.  If only
synchrotron photons contribute to the TeV $\gamma$-ray component via the
SSC mechanism and the injected electron density changes as in the case
in Mrk~421, the synchrotron flux will be proportional to the electron
density ($F_{\rm Sync}\propto N_{\rm e}$), and the IC TeV $\gamma$-ray
flux will be proportional to both the electron density and synchrotron
photon flux as $F_{\rm SSC}\propto N_{\rm e} \cdot F_{\rm Sync}$. In
this case, the correlation of $F_{\rm SSC}\propto F_{\rm Sync}^{2}$ is
expected.  On the other hand, if the seed photons of the IC scattering
process are fed by external photons from the accretion disk
\cite{Dermer1992} or from the central region of the AGN
\cite{Sikora1994}, then the TeV $\gamma$-ray/keV X-ray correlation may
come down to a linear correlation.

On the other hand, if we examine the correlation closely, some intense
TeV $\gamma$-ray flares seem to have occurred without large X-ray
flares as seen in the flare phases except for Term 3 (meshed circles
in Fig.\ 9).  These results may suggest that the physical parameters
such as magnetic field strength, injected electron density and its
spectral index, seed photon density, etc. \cite{Petry2000}, in the
shock region of the AGN jet differ flare by flare.

While no definite conclusion can be obtained regarding the
correlation discussed above based on the present experiment, it is very
important to continue the long-term simultaneous multi-wavelength
observation of Mrk~421 flares and to accumulate sufficient
experimental data in order to understand the mechanism of TeV
$\gamma$-ray emissions from Mrk~421 and Mrk~501.  It should be stressed
that a true long-term simultaneous TeV $\gamma$-ray/keV X-ray
observation is available by means of only the combination of an X-ray satellite experiment
and a wide field-of-view air shower experiment.

\subsection{ENERGY SPECTRUM OF FLARING TeV $\gamma$-RAYS}

In order to determine the energy spectrum of flaring $\gamma$-rays from
Mrk~421, we divided the Tibet~III dataset into 3 terms according to
the period that the ASM/RXTE satellite recorded more than 1.0 ASM
units (counts/sec) averaged over 30 days, which is shown by three gray
areas in Figure~7. We also calculated the detector response of the
Tibet~III array based on the full MC simulation. For this,
$\gamma$-rays from Mrk~421 are simulated, assuming a differential
power-law spectrum of $E^{-\beta}$, where $\beta$ varies from 2.6
to 6.0 and taking into account the diurnal motion of Mrk~421 in the
sky. Air shower events are uniformly thrown within a circle with a
radius of 300~m whose center is positioned on the center of the array. This
radius is sufficient to collect all $\gamma$-ray events which are
actually triggered in our array.  Using the calculated effective area,
the excess event rate, live time, and the relation between
$\sum\rho_{\rm FT}$ and the primary $\gamma$-ray energy, we can calculate
the differential energy spectrum of $\gamma$-rays from Mrk~421.  In
the present work, the energy points indicate the log-scale mean of energies in
each $\sum\rho_{\rm FT}$ bin defined as follows: 
10 $\times$ $10^{n / 4}$ $<$ $\sum\rho_{\rm FT}$ 
$\leq$ 10 $\times$ $10^{(n+1) / 4}$ ($n = 0, 1, 2, 3$), and 
100 $\times$ $10^{n / 3}$ $<$ $\sum\rho_{\rm FT}$ 
$\leq$ 100 $\times$ $10^{(n+1) / 3}$ ($n = 0, 1, 2$), 
where the lowest energy bin is dropped off in the analysis
because the trigger efficiency is estimated to be very low ($<$1\%).
Thus, the total number of energy bins available in the analysis is
six.

We obtained the differential energy spectra for the 3 terms and their
power indices($\beta$) are summarized in Table 1.  Figure~10 shows the
differential energy spectrum observed in the most active phase in 2001
(Term 3), with a spectral index of $-$3.24$\pm$0.69 at energy range between 1~TeV and 5~TeV,
together with those obtained by the IACT experiments during 2001.
Figure~10 (inset) shows the $\sum\rho_{\rm FT}$ spectrum of
observed $\gamma$-ray events from Mrk~421, together with those from
the MC simulation assuming spectral indices of $-$3.24 (solid line) and of
$-$2.8 (dashed line), respectively.
Note that the absolute flux cannot be discussed in terms of Figure~10
because our observation period did not overlap perfectly those of IACTs, 
and that the absolute flux is estimated in all experiments at an error of
approximately $\pm$20\% which originates from
the uncertainty of the absolute energy scale.

It is considered that the energy spectrum may become steep or break at
high energies due to several reasons such as the interaction of
$\gamma$-rays with the infrared photon field in the
intergalactic space \cite{DeJager1994}, the photon-photon pair
attenuation near the source \cite{Dermer1994}, and the Klein-Nishina
scattering cross section limit \cite{Hillas1999}.  A cutoff energy
($E_c$) of the spectrum may be estimated by fitting a spectrum form of
$E^{-\beta}\exp(-E/E_c)$ to the observed data. For example, the HEGRA
group \cite{Aharonian2002} estimated a cutoff energy to be
3.6$^{+0.4}_{-0.3 {\rm stat}}$$^{+0.9}_{-0.8 {\rm syst}}$ TeV and
6.2$\pm 0.4_{\rm stat}$$^{+2.9}_{-1.5 {\rm syst}}$ TeV for Mrk~421
and Mrk~501, respectively, and considered that this difference may be
attributed to some intrinsic difference between $\gamma$-ray emission
mechanisms.  On the other hand, the Whipple group \cite{Krennrich2001}
reported a cutoff energy to be 
4.3$\pm 0.3_{\rm stat}$$^{+1.7}_{-1.4{\rm syst}}$ TeV 
and 4.6$\pm0.8$$_{\rm stat}$ TeV for Mrk~421 and
Mrk~501, respectively. They argue that the same cutoff energy obtained
for both sources having almost the same redshift may be due to
the absorption by the infrared photon field in the universe. Our observed
spectrum for Mrk~421 is not inconsistent with those from the HEGRA,
Whipple, and CANGAROO groups within statistical and systematic errors.

\section{SUMMARY}

Mrk~421 was in an active phase during the period between the year 2000
and 2001, showing strong and frequent flaring.  During this flaring
period, the Tibet~III array successfully monitored the sky region with
$\sim$2-steradian solid angle.  This constant observation is beyond
reach of IACTs which can observe the sky only on clear moonless
nights.  The stability of the array operation can be well checked by
continuously observing the Moon's shadow and the event rate of air shower events.
Using this array, we detected multi-TeV flaring
$\gamma$-rays from Mrk~421 at a significance level of 5.1~$\sigma$
and found a positive flux correlation between the keV and TeV energy
regions.  The Tibet~III and the ASM/RXTE data seem to favor a
linear correlation rather than a quadratic one, although 
these data have not yet expressed sufficient statistical significance
for this finding to be certain.  Here, it should
be stressed that the Tibet~III array succeeded in the first
observation of long-term correlation between satellite keV X-ray and
TeV $\gamma$-ray data based on simultaneous observation. The observed
energy spectrum with a power index of $-$3.24$\pm$0.69 is not
inconsistent with those measured by IACTs within statistical
and systematic errors.

The area of the Tibet~III array was further enlarged up to 36,900
m$^2$ by adding 200 counters in the late fall of 2002 and this new
array has been successfully operating since then, triggering
air shower events at a rate of 1,450~Hz with a dead time of about 10\%. 
With the advent of the full-scale Tibet~III array, long-term
observation of TeV $\gamma$-rays from Mrk~421 or Mrk~501 together with
other wavelength data may lead in the near future to a deeper understanding of the
$\gamma$-ray emission mechanism. Moreover, the
successful observation of $\gamma$-rays from the Crab~Nebula, Mrk~421, and
Mrk~501 by the Tibet air shower experiment demonstrates that we are
now ready for detecting unknown stable/transient TeV $\gamma$-ray
point sources through an all-sky survey.

\acknowledgments

  This work is supported in part by Grants-in-Aid for Scientific
Research on Priority Area, for Scientific Research and also for
International Science Research from the Ministry of Education,
Science, Sports and Culture in Japan, and for International Science
Research from the Committee of the Natural Science Foundation and the
Chinese Academy of Sciences in China.


\begin{figure}
\epsscale{0.7}
\plotone{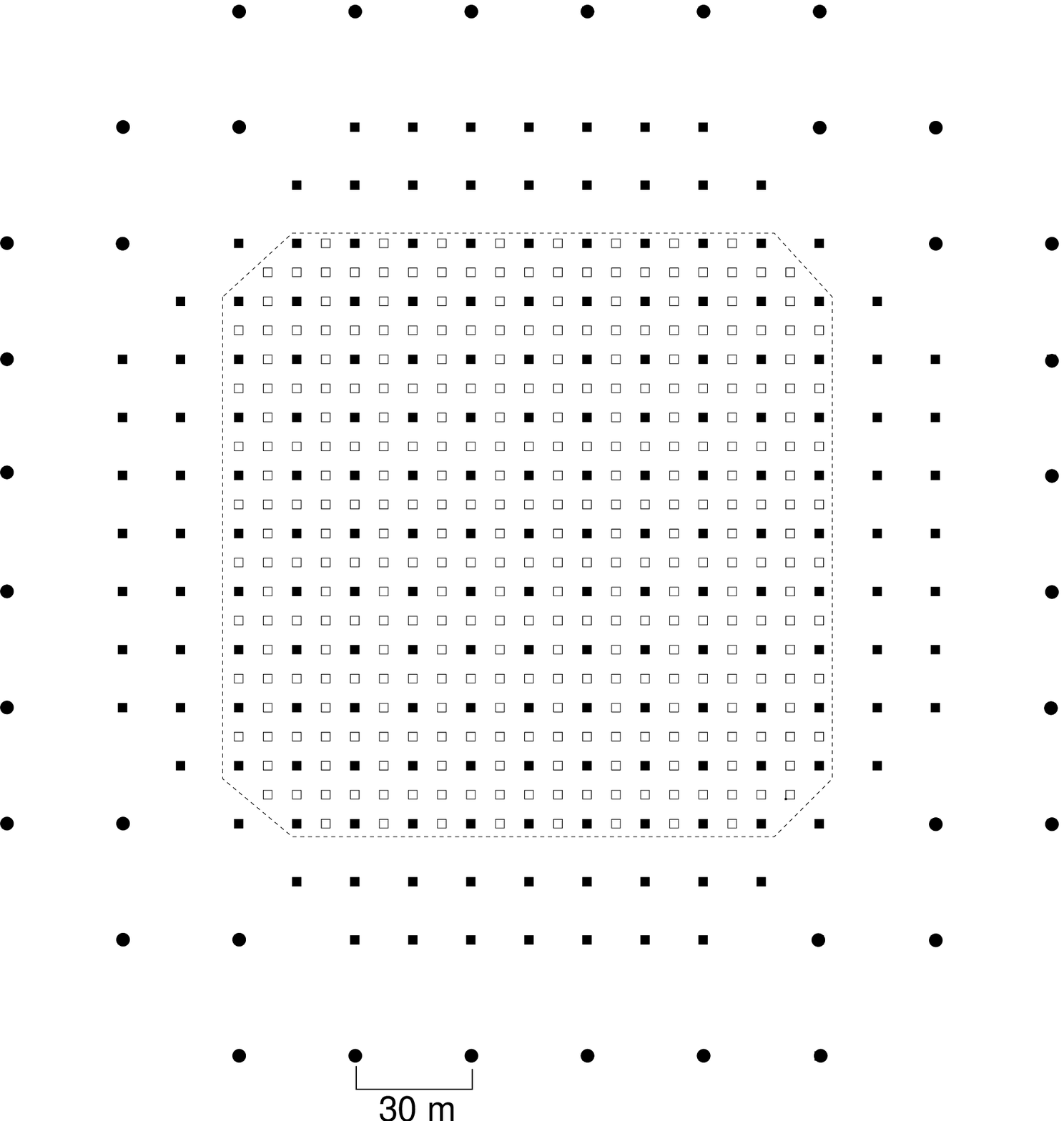}
\caption{ Schematical view of the Tibet~III array operating at
Yangbajing.  Open squares: FT-detectors; filled squares: FT-detectors
with a wide dynamic range PMT; filled circles: density detectors with
a wide dynamic range PMT. We select air shower events whose cores
are located within the detector matrix enclosed by the dashed line.
\label{fig1}}
\end{figure}

\clearpage

\begin{figure}
\epsscale{0.9}
\plotone{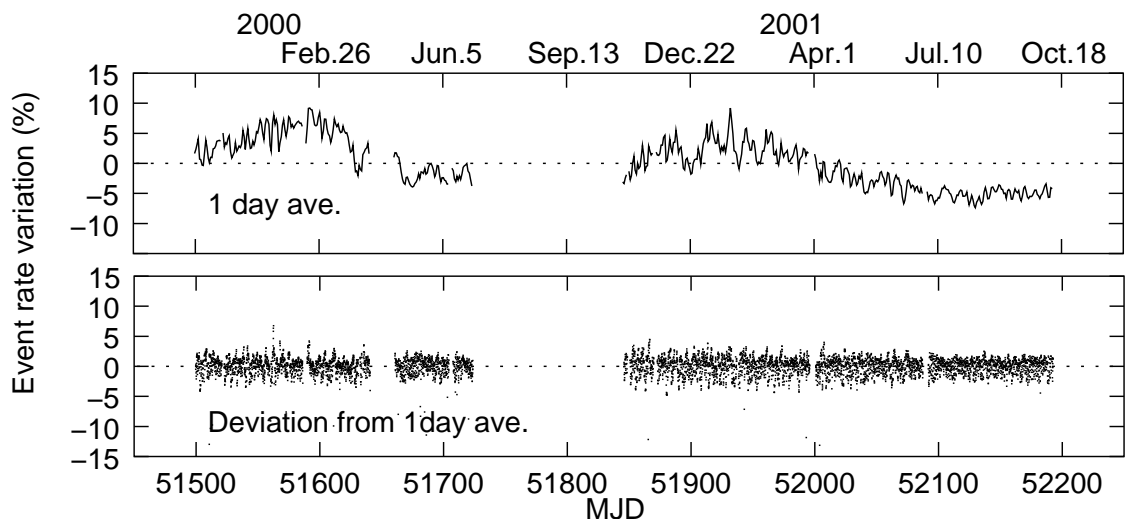}
\caption{ Tibet~III long-term event rate stability, where 
corrections for the atmospheric pressure and temperature effects are not
made.  The upper panel shows the 1-day average event rate variation
($\pm$5\%).  The lower panel shows the 2-hour average day/night
event rate variation ($\pm$2\%) after subtracting the 1-day average
event rate.
\label{fig2}}
\end{figure}

\begin{figure}
\epsscale{0.9}
\plotone{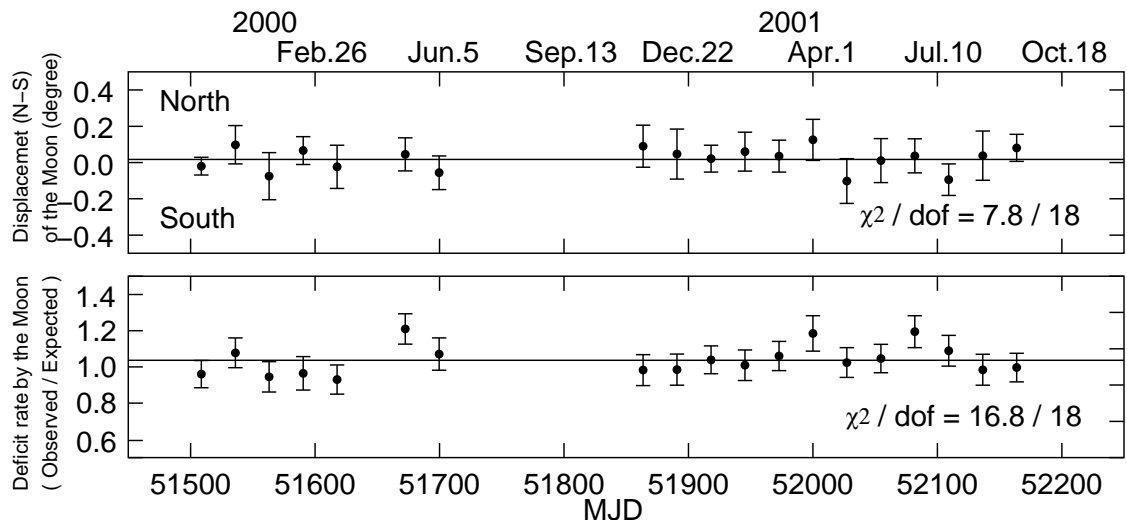}
\caption{ Upper panel shows the displacement of the Moon's shadow
center in the north-south direction.  The lower panel shows the
long-term stability of the ratio of the observed deficit counts to the
expected deficit counts within a circle having an angular radius of
$0.^{\circ}9$.
\label{fig3}}
\end{figure}

\clearpage

\begin{figure}
\epsscale{0.5}
\plotone{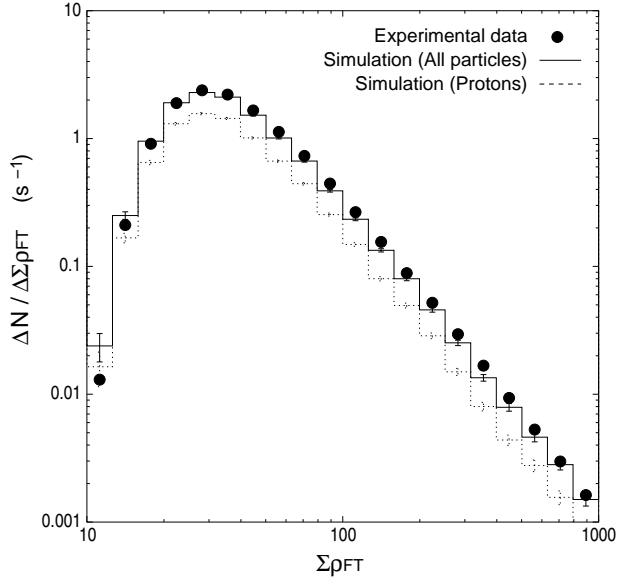}
\caption{ $\sum\rho_{\rm FT}$ spectrum. Closed circles denote
experimental data.  Solid histograms denote the simulation assuming a
primary cosmic-ray flux model based on directly observed data,
while the dashed histograms represent the events induced by protons.
\label{fig4}}
\end{figure}

\begin{figure}
\epsscale{0.48}
\plotone{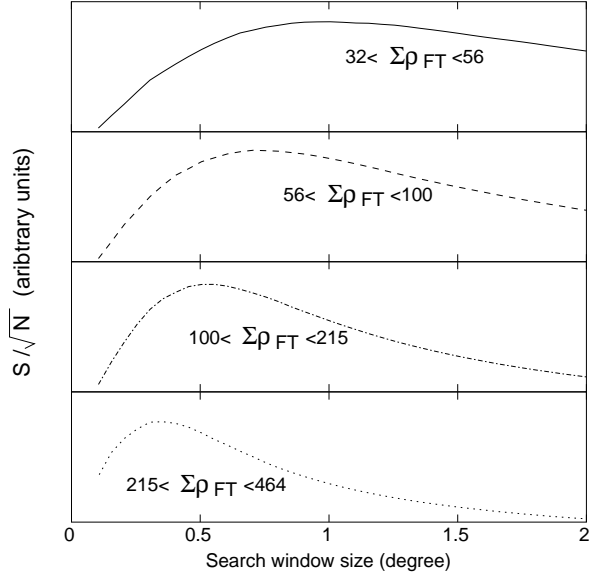}
\caption{ 
Typical $S/\sqrt{N}$ curves as a function of search window size
for each $\sum\rho_{\rm FT}$ bin.
Solid curve: 32~$<$~$\sum\rho_{\rm FT}$~$\leq$~56;
Dashed curve: 56~$<$~$\sum\rho_{\rm FT}$~$\leq$~100;
Dash-dot curve: 100~$<$~$\sum\rho_{\rm FT}$~$\leq$~215.
Dotted curve: 215~$<$~$\sum\rho_{\rm FT}$~$\leq$~464.
We obtain the optimal search window radius ($6.9 / \sqrt{ \sum\rho_{\rm FT} }$
degree) as a function of $\sum\rho_{\rm FT}$ which maximizes the
$S/\sqrt{N}$ ratio.
\label{fig5}}
\end{figure}

\begin{figure}
\epsscale{0.6}
\plotone{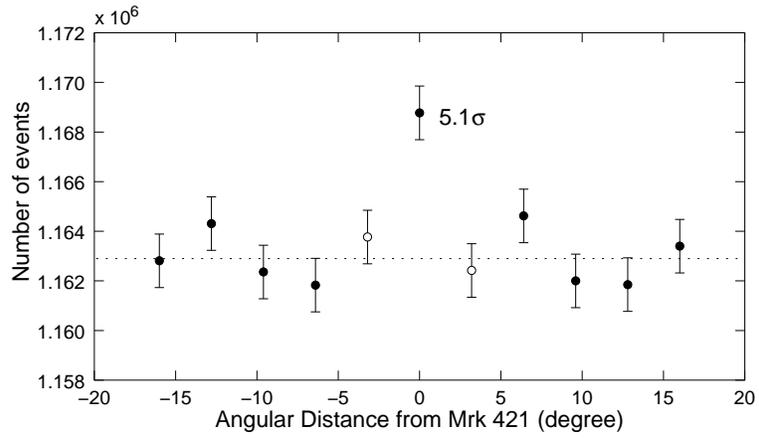}
\caption{ Number of observed air shower events with energies $>$ 1 TeV
after event reduction during an observation time of 456.8 live
days in 2000 and 2001 as a function of angular distance from Mrk~421
in the azimuthal direction.
\label{fig6}}
\end{figure}

\clearpage

\begin{figure}
\epsscale{0.9}
\plotone{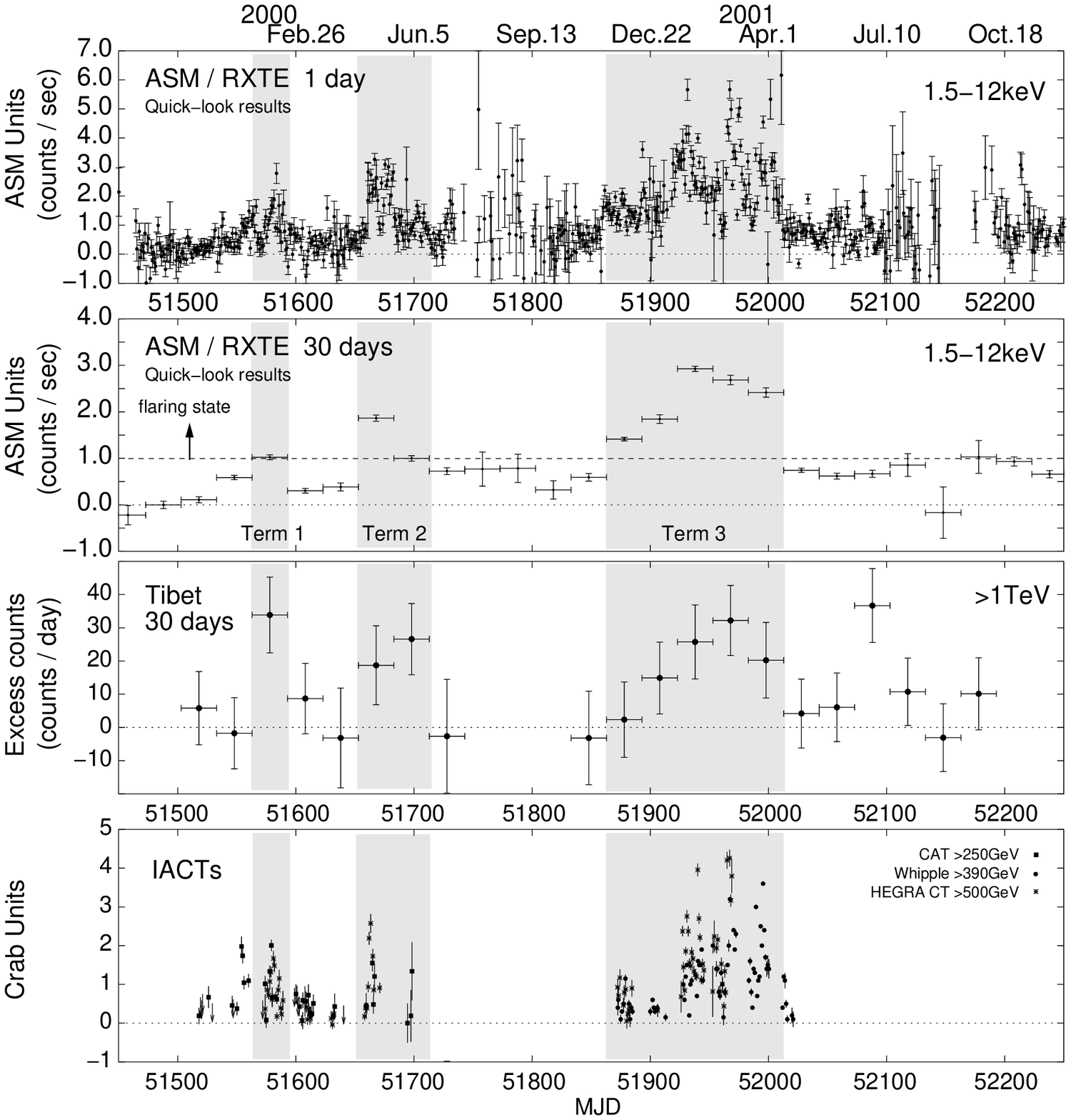}
\caption{ Daily excess event rate averaged over 30 days around the
Mrk~421 flaring period in the years 2000 and 2001, together with 
ASM/RXTE X-ray satellite observation \citep{Levine1996,RXTEhtml}. Also
shown are the observations by IACTs; the Whipple \cite{Holder2001}, HEGRA
\cite{Kohnle2001}, CAT \cite{Piron2001}.  The Tibet~III dataset is
divided into three active terms according to the flaring states of
Mrk~421. In each term, the ASM/RXTE satellite recorded more than 1.0
ASM units (counts/sec) averaged over 30 days, as is indicated by three
gray areas.
\label{fig7}}
\end{figure}

\clearpage

\begin{figure}
\epsscale{0.6}
\plotone{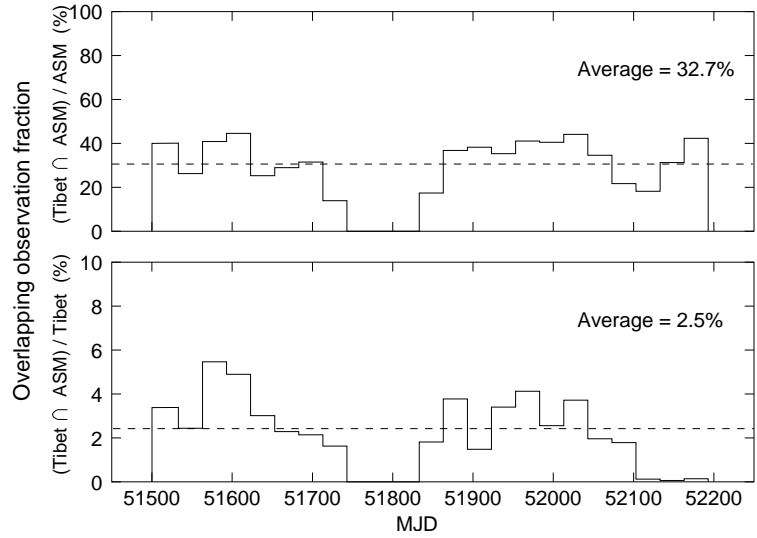}
\caption{ Time variation of fractions of Mrk~421 observation times
overlapping between the Tibet~III array ($T_{\rm Tibet}$) and the
X-ray satellite ASM/RXTE ($T_{\rm ASM}$), averaged over one month.
The upper histograms show 
($T_{\rm Tibet}$ $\cap$ $T_{\rm ASM}$) / $T_{\rm ASM}$, 
while the lower histograms show 
($T_{\rm Tibet}$ $\cap$ $T_{\rm ASM}$) / $T_{\rm Tibet}$.
\label{fig8}}
\end{figure}

\clearpage

\begin{figure}
\epsscale{0.6}
\plotone{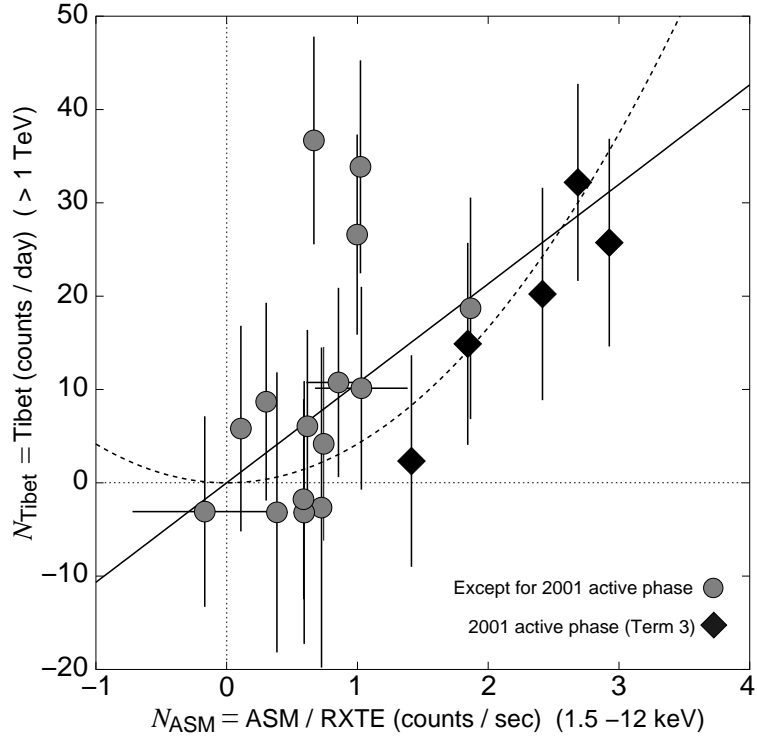}
\caption{ Correlation between the Tibet~III excess counts and ASM/RXTE
count rates.  A quadratic fit to the data points (meshed circles and
filled diamonds) yields 
$N_{\rm Tibet}$~=~(4.16$\pm$0.81) $N_{\rm ASM}^{2}$, 
$\chi^{2} / d.o.f. = 24.6 / 19$ shown by a dashed curve,
while a linear fit to the data gives 
$N_{\rm Tibet}$~=~(10.66$\pm$1.83) $N_{\rm ASM}$,
$\chi^{2} / d.o.f. = 17.6 / 19$ shown by a solid line.  
The filled diamonds show a correlation at the most
active phase in 2001 (Term 3 in Fig.\ 7), while the meshed circles
show a correlation at the flare phases except for Term 3.
\label{fig9}}
\end{figure}

\begin{figure}
\epsscale{0.6}
\plotone{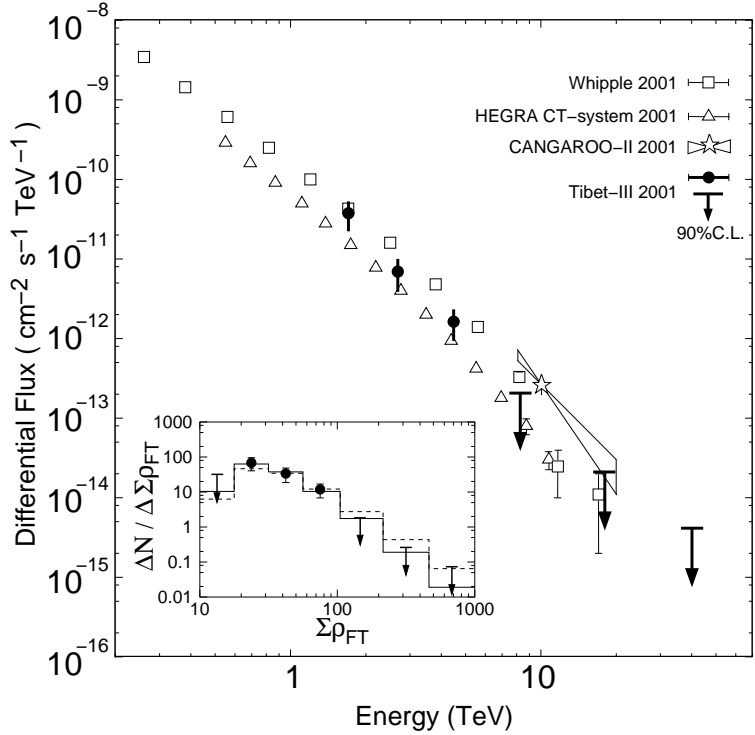}
\caption{ Differential energy spectrum of $\gamma$-rays from Mrk~421
at the most active phase in 2001 (Term 3) observed by the Tibet~III
array, together with those obtained by the IACTs, i.e., the Whipple
\cite{Krennrich2001}, HEGRA \cite{Kohnle2001}, and CANGAROO
\cite{Okumura2002} during mostly overlapping periods.  The
Tibet~III upper limits are given at the 90\% confidence level,
according to a statistical subscription \cite{Helene1983}.  The inset
figure shows the $\sum\rho_{\rm FT}$ spectrum of the observed
$\gamma$-ray events from Mrk~421, together with the simulations
assuming a spectral index of $-$3.24 (solid histograms) and $-$2.8
(dashed histograms), respectively.
\label{fig10}}
\end{figure}

\clearpage

\begin{deluxetable}{llcccc}
\tabletypesize{\scriptsize} 
\tablecaption{ Best fit differential
spectral index of $\gamma$-rays from Mrk~421 detected by the
Tibet~III array in the three active phases as defined in Figure~7 in
the years 2000 and 2001.
\label{tbl-1}}
\tablewidth{0pt}
\tablehead{
\colhead{Term Name\tablenotemark{a}}& \colhead{Spectral Index} &
\colhead{Energy range}  &{$\chi^2$}  & \colhead{$d.o.f.$} \\
\colhead{}& \colhead{} &
\colhead{(TeV)}  &{}  & \colhead{}
}
\startdata
Term1+2+3&-3.77$\pm$0.50& 1 - 5 &0.8&1\\
Term1+2  &-4.52$\pm$0.80& 1 - 5 &0.5&1\\
Term3    &-3.24$\pm$0.69& 1 - 5 &0.2&1\\
\enddata
\tablenotetext{a}{
MJD of Term~1, 2 and 3 ranges 51563-51593, 51653-51713 and 51863-52023, 
respectively.
}
\end{deluxetable}

\clearpage

\end{document}